# Characterization of quantum states based on creation complexity

Zixuan Hu and Sabre Kais*

*Department of Chemistry, Department of Physics, and Birck Nanotechnology Center, Purdue University, West Lafayette, IN 47907, United States*
*Email:* kais@purdue.edu

The creation complexity of a quantum state is the minimum number of elementary gates required to create it from a basic initial state. The creation complexity of quantum states is closely related to the complexity of quantum circuits, which is crucial in developing efficient quantum algorithms that can outperform classical algorithms. A major question unanswered so far is what quantum states can be created with a number of elementary gates that scales polynomially with the number of qubits. In this work we first show for an entirely general quantum state it is exponentially hard (requires a number of steps that scales exponentially with the number of qubits) to determine if the creation complexity is polynomial. We then show it is possible for a large class of quantum states with polynomial creation complexity to have common coefficient features such that given any candidate quantum state we can design an efficient coefficient sampling procedure to determine if it belongs to the class or not with arbitrarily high success probability. Consequently partial knowledge of a quantum state's creation complexity is obtained, which can be useful for designing quantum circuits and algorithms involving such a state.

## I. Introduction

Quantum computing has seen enormous progress in both the theoretical and the experimental fronts[1-6]. From the early proposals of the phase estimation algorithm[7], the Shor's factorization algorithm[8] and the Harrow-Hassidim-Lloyd algorithm for linear systems[9], to the more recent ones like the variational quantum eigensolver[10], the quantum machine learning algorithms[11,12], and quantum algorithm for open quantum dynamics[13,14], the potential of quantum algorithms to outperform their classical counterparts in numerous tasks become increasingly realistic with the rapid development of quantum computing hardware[15-18]. However, the design of quantum algorithms so far remains an accidental process because there is no systematic way to look for algorithms that scale efficiently. For any quantum algorithm, the potential of outperforming classical algorithms lies in the efficient scaling of the quantum circuit complexity. In general if the complexity of a quantum circuit – as measured by the number of elementary gates used – scales polynomially with the number of qubits involved, it is efficient compared to the classical algorithm for the same purpose. Since any quantum circuit is essentially a sequence of unitary gates operating on an input state and producing an output state, we can relate the complexity of creating the output state to the complexity of the quantum circuit. Indeed if a quantum circuit involves a critical step at which a critical intermediate quantum state has a fixed form with a given creation complexity, then the complexity of the circuit can be evaluated by the creation complexity of this intermediate state. It is well known that the overwhelming majority of all quantum states have creation complexities that scale exponentially with the number of qubits[19] – i.e. the creation complexity is exponential. The problem of designing efficient quantum circuits therefore may involve identifying those quantum states with polynomial creation complexity. Now for a given known quantum state, can we determine if its creation complexity is polynomial? If the answer is yes, does the determination process itself have a polynomial complexity? These are two open questions (so far as we know) examined by this study. These questions are important because if there is an efficient and systematic way to determine if a quantum state is polynomial (having polynomial creation complexity), we can then design quantum algorithms using only the polynomial states such that polynomial complexity of the circuit is guaranteed. This will guide our search for new efficient algorithms by significantly reducing the search space.

In the following we first prove that for an entirely general quantum state it is exponentially hard to determine if the creation complexity is polynomial. Next we show that this apparently discouraging result does not prevent us from obtaining partial knowledge of the creation complexity of a given candidate state. In particular, we can define a proper subset of all the polynomial states such that all states in the subset have some simple characteristics that can be identified efficiently. Given a candidate state, we will determine if it has the characteristics or not: if yes it belongs to the subset and is polynomial; if not it does not belong to the subset and its creation complexity is undecided because the subset does not contain all the polynomial states. As the negative result leaves the candidate state undecided, it is desirable to have a very general subset such that it contains states with great generality and complexity. In this work we identify such a subset that includes those quantum states with great generality and the maximum complexity allowed by the number of qubits. We then analyze the characteristics of our subset and show how an arbitrary state can be transformed into a form where these characteristics may be found. Finally we propose a method to determine if an arbitrary candidate state belongs to the subset, and if yes design an efficient quantum circuit to create the state.

## II.    Theory

*1. The cost of determining the creation complexity of a general state.* To determine the creation complexity of a general quantum state we need to first define the basic initial state and the set of elementary gates. In this work the basic initial state is naturally defined as the all zero n-qubit product state $|0\rangle^{\otimes n}$, and the elementary gates include all 1-qubit unitaries and the 2-qubit CNOT. First we note the effect of the 1-qubit unitaries is to rotate the qubit in the 2-dimensional Hilbert space, and the effect of the CNOT gates is to entangle the two qubits involved in the gate. Without the CNOT gates, the states we can create with only 1-qubit unitaries have the maximum creation complexity of $n$ with one unitary per qubit, and any additional application of 1-qubit unitaries is redundant. On the other hand, without the 1-qubit unitaries, the CNOT gates can only shuffle the coefficients of the starting state vector around, and cannot create new coefficient values. Therefore we can say the 1-qubit unitaries and CNOT gates play complementary roles, and can be combined to define a new elementary set containing only 2-qubit controlled-unitary gates or $C(U)$. It is easy to verify that all possible 1-qubit unitaries and the 2-qubit CNOT gates can be expressed in $C(U)$'s, thus that the set of all $C(U)$'s is universal. The universal set of $C(U)$'s is equivalent to the universal set of 1-qubit unitaries and CNOT gates in the sense that a state with polynomial creation complexity in one universal set always has the same creation complexity in another universal set. Working only with the $C(U)$'s allows us to easily evaluate the cost of determining the creation complexity of a general quantum state.

If we define a general quantum state by the minimum sequence of $C(U)$ gates required to create it from the initial $|0\rangle^{\otimes n}$, we have a map between a tuple of $\left(C(U_1), C(U_2), ..., C(U_N)\right)$ and a quantum state, where the tuple represents an ordered sequence of $C(U_i)$'s. All quantum states with polynomial creation complexity correspond to all the tuples of $C(U_i)$'s with the lengths smaller than $cn^k$ for some constant $c$ and $k$ such that $cn^k$ is overwhelmingly smaller than $2^n$, i.e. $2^n - cn^k \approx 2^n$. Since each $C(U_i)$ has $2^2 = 4$ free real parameters, the subspace formed by all tuples of $C(U_i)$'s with length smaller than $cn^k$ has the dimension of at most $4cn^k$. On the other hand a general n-qubit state has $2^{n+1} - 2$ free real parameters ($2^n - 1$ free complex parameters) for its state vector. To determine if a candidate state is in the set of states with polynomial creation complexity, we need to tell if a vector in an $\mathbf{R}^{2^{n+1}-2}$ space belongs to the subspace of $\mathbf{R}^{4cn^k}$, and this in general requires that $2^{n+1} - 2 - 4cn^k$ conditions be checked (the co-dimension of $\mathbf{R}^{4cn^k}$ in $\mathbf{R}^{2^{n+1}-2}$ is $2^{n+1} - 2 - 4cn^k$). Because $2^n - cn^k \approx 2^n$, $2^{n+1} - 2 - 4cn^k$ is also a number exponential in $n$, so that not only it is difficult to identify all these conditions, but it is also difficult to check these conditions one by one. *We therefore conclude that for an entirely general quantum state it is exponentially hard to determine if the creation complexity is polynomial.* This however does not mean that given any candidate state we can never obtain any knowledge of its creation complexity. Out of all the states with polynomial creation complexity, there exist subsets of states with characteristics easy to determine such that given a candidate state we can determine with polynomial steps if it belongs to the subset. If a state belongs to the subset, then its creation complexity is polynomial, otherwise its creation complexity is unknown. This partial knowledge of a state's creation complexity can be used to guide the development of quantum algorithms if the subset is big enough to include many quantum states of potential interest. A simple example of such a subset includes all the states with fewer than $cn^k$ number of non-zero elements in its state vector. As the initial state $|0\rangle^{\otimes n}$'s state vector has only 1 non-zero element at the first entry, if we can introduce one new non-zero element in any other entry with polynomial number of steps, e.g. $dn^l$, then the state vector with fewer than $cn^k$ number of non-zero elements can be created with $cn^k \cdot dn^l = cdn^{k+l}$ number of steps – this may not be the optimal procedure but good enough as it is polynomial. This is indeed possible by a 2-level unitary matrix involving the first entry and any other entry that we want to add[19]. Checking if a given candidate state is in this subset is easy because the conditions are identified as some simple characteristics of the coefficients: having $cn^k$ non-zero elements. Although requiring $2^n$ steps, it is already much easier to check one-by-one if each entry of the state vector is zero than to identify some complex conditions for the general set of states with polynomial creation complexity. Furthermore, as shown below in Procedure 1 for a more general subset, once the conditions are identified with some characteristics of the coefficients, it is possible to design a polynomial procedure to check a candidate state's coefficients and

determine if it belongs to the subset with arbitrarily high probability (no longer a deterministic test). The subset of states with fewer than $cn^k$ number of non-zero elements in its state vector can therefore provide partial knowledge for a candidate state. However, the knowledge we can obtain is very limited because this subset is a very special one. As the overwhelming majority of the entries in the state vector are zero, the possible states are confined to low-dimensional subspaces. In particular, this subset cannot have states with a large number of qubits entangled with each other. In the following, we define a much more general subset of states with polynomial creation complexity such that states with arbitrary number of non-zero elements and maximum entanglement (all qubits are entangled) are included. This will provide much more knowledge for the creation complexity of a candidate state.

*2. The standard state and the polynomial standard state.* We have seen that the effect of the 1-qubit unitaries is to rotate the qubit in the 2-dimensional Hilbert space, and the effect of the CNOT is to entangle the two qubits involved in the gate. Next we will try to estimate the minimum number of steps to create a quantum state by applying the effects of the elementary gates without redundancy. Take a 2-qubit state for example, the most general form of the state is given by the Schmidt decomposition:

$$\phi^{(2)} = a_1 \left| u_1 v_1 \right\rangle + a_2 \left| u_2 v_2 \right\rangle \tag{1}$$

where the $(2)$ on $\phi^{(2)}$ denotes a 2-qubit entangled state, $\left| u_1 \right\rangle$ and $\left| u_2 \right\rangle$ are orthogonal, $\left| v_1 \right\rangle$ and $\left| v_2 \right\rangle$ are orthogonal, $a_1$ and $a_2$ are a pair of complex coefficients satisfying $\left| a_1 \right|^2 + \left| a_2 \right|^2 = 1$. To create $\phi^{(2)}$ with minimal steps we can first apply a unitary $U_2$ on the second qubit $q_2$ to produce $a_1 \left| 0 \right\rangle_2 + a_2 \left| 1 \right\rangle_2$, then apply $\text{CNOT}_{2 \to 1}$ ( $2 \to 1$ means using $q_2$ to control $q_1$ ) to produce $a_1 \left| 00 \right\rangle + a_2 \left| 11 \right\rangle$, then apply two more unitary gates on $q_1$ and $q_2$ respectively to produce $\phi^{(2)} = a_1 \left| u_1 v_1 \right\rangle + a_2 \left| u_2 v_2 \right\rangle$. We remark that this procedure applies the effects of the elementary gates in a very efficiently way: first $U_2$ paired with $\text{CNOT}_{2 \to 1}$ to create the entanglement, then two unitaries to rotate the qubits into the final forms. For a general $\phi^{(2)}$ these are the minimum operations required. Can we apply more gate operations on $\phi^{(2)}$ to make the state more complex? The answer is no, because any additional gates will only transform $\phi^{(2)}$ into another Schmidt form $\phi'^{(2)} = a'_1 \left| u'_1 v'_1 \right\rangle + a'_2 \left| u'_2 v'_2 \right\rangle$ and the same procedure as above with modified parameters can produce the new state. A general $\phi^{(2)} = a_1 \left| u_1 v_1 \right\rangle + a_2 \left| u_2 v_2 \right\rangle$ is therefore the 2-qubit state with the maximum creation complexity. Generalizing to an n-qubit state there is a most general form $\phi^{(n)}$ with the maximum creation complexity, and to increase the creation complexity further we need to add more qubits into the entanglement and increase the number $n$. The explicit form of $\phi^{(n)}$ can be obtained by Schmidt-decomposing it top-down iteratively as the following:

1. decompose $\phi^{(n)}$ with respect to one qubit :

$$\phi^{(n)} = C_1 \phi_1^{(n-1)} \left| u_1 \right\rangle + C_2 \phi_2^{(n-1)} \left| u_2 \right\rangle$$

2. decompose $\phi_1^{(n-1)}$ and $\phi_2^{(n-1)}$ with respect to another qubit :

$$\phi^{(n)} = C_1 \left[ D_{11}\phi_{11}^{(n-2)} \left| v_{11} \right\rangle + D_{12}\phi_{12}^{(n-2)} \left| v_{12} \right\rangle \right] \left| u_1 \right\rangle$$
$$+ C_2 \left[ D_{21}\phi_{21}^{(n-2)} \left| v_{21} \right\rangle + D_{22}\phi_{22}^{(n-2)} \left| v_{22} \right\rangle \right] \left| u_2 \right\rangle$$

3. decompose $\phi_{11}^{(n-2)}$, $\phi_{12}^{(n-2)}$, $\phi_{21}^{(n-2)}$ and $\phi_{22}^{(n-2)}$ with respect to another qubit:

$$\phi^{(n)} = C_1 \left[ \begin{array}{l} D_{11}\left( E_{111}\phi_{111}^{(n-3)} \left| w_{111} \right\rangle + E_{112}\phi_{112}^{(n-3)} \left| w_{112} \right\rangle \right) \left| v_{11} \right\rangle \\ + D_{12}\left( E_{121}\phi_{121}^{(n-3)} \left| w_{121} \right\rangle + E_{122}\phi_{122}^{(n-3)} \left| w_{122} \right\rangle \right) \left| v_{12} \right\rangle \end{array} \right] \left| u_1 \right\rangle$$
$$+ C_2 \left[ \begin{array}{l} D_{21}\left( E_{211}\phi_{211}^{(n-3)} \left| w_{211} \right\rangle + E_{212}\phi_{212}^{(n-3)} \left| w_{212} \right\rangle \right) \left| v_{21} \right\rangle \\ + D_{22}\left( E_{221}\phi_{221}^{(n-3)} \left| w_{221} \right\rangle + E_{222}\phi_{222}^{(n-3)} \left| w_{222} \right\rangle \right) \left| v_{22} \right\rangle \end{array} \right] \left| u_2 \right\rangle$$

4. decompose the eight $\phi_{ijk}^{(n-3)}$ states with respect to another qubit ...

... this continues iteratively to the $\phi^{(2)}$ level $\qquad$ (2)

In Eq.(2) if we continue the iteration to the lowest level of 2-qubit states, there are at most $2^{n-2}$ $\phi^{(2)}$ terms and from the bottom-up it looks like:

$$\phi^{(n)} = \left\{ \begin{array}{l} \left\{ \begin{array}{l} \left[ \begin{array}{l} \left( a_1 \left| u_1 v_1 \right\rangle + a_2 \left| u_2 v_2 \right\rangle \right) b_1 \left| w_1 \right\rangle \\ + \left( a_3 \left| u_3 v_3 \right\rangle + a_4 \left| u_4 v_4 \right\rangle \right) b_2 \left| w_2 \right\rangle \end{array} \right] c_1 \left| x_1 \right\rangle \\ + \left[ \begin{array}{l} \left( a_5 \left| u_5 v_5 \right\rangle + a_6 \left| u_6 v_6 \right\rangle \right) b_3 \left| w_3 \right\rangle \\ + \left( a_7 \left| u_7 v_7 \right\rangle + a_8 \left| u_8 v_8 \right\rangle \right) b_4 \left| w_4 \right\rangle \end{array} \right] c_2 \left| x_2 \right\rangle \end{array} \right\} d_1 \left| y_1 \right\rangle \\ + \left\{ \qquad \qquad \ldots \ldots \qquad \qquad \right\} d_2 \left| y_2 \right\rangle \end{array} \right\} \ldots \qquad (3)$$

where we have used $a_i$'s to represent the coefficients $C_i^{(2)}$ on the $\phi^{(2)}$ level, $b_i$'s to represent the coefficients $C_i^{(3)}$ on the $\phi^{(3)}$ level, $c_i$'s to represent the coefficients $C_i^{(4)}$ on the $\phi^{(4)}$ level, and so on. From Eq.(3), we see the most general $\phi^{(n)}$ contains $2^{n-2}$ unique pairs of the coefficients $\{a_{2i-1}, a_{2i}\}$ on the $\phi^{(2)}$ level, $2^{n-3}$ unique pairs of the coefficients $\{b_{2i-1}, b_{2i}\}$ on the $\phi^{(3)}$ level, and the trend continues for the coefficients on each level. This means if we want to specify all the unique pairs of coefficients for a most complex $\phi^{(n)}$, the cost is already exponential in $n$, while we have not yet considered the cost to specify the qubit states on each level such as the $\left| u_i \right\rangle$'s, $|v_i\rangle$'s and $|w_i\rangle$'s. Thus we can see the problem of quantum state creation is exponentially hard, and the goal of the following is to characterize a general subset of states with polynomial creation complexity such that it can provide partial knowledge of the creation complexity of a candidate state. First we note that the procedure described above for $\phi^{(2)}$ can be easily generalized to create a special $\phi^{(n)}$ with only polynomial steps:

Start of the procedure with the initial state $|0\rangle^{\otimes n}$

1. $U_2$ on $q_2$ followed by $\text{CNOT}_{2\to 1}$ :
$$a_1|00\rangle_{12} + a_2|11\rangle_{12}$$

2. $U_3$ on $q_3$ followed by $\text{CNOT}_{3\to 2}$ :
$$\left(a_1|00\rangle_{12} + a_2|11\rangle_{12}\right)b_1|0\rangle_3 + \left(a_1|01\rangle_{12} + a_2|10\rangle_{12}\right)b_2|1\rangle_3$$

3. $U_4$ on $q_4$ followed by $\text{CNOT}_{4\to 3}$ :
$$\left[\left(a_1|00\rangle_{12} + a_2|11\rangle_{12}\right)b_1|0\rangle_3 + \left(a_1|01\rangle_{12} + a_2|10\rangle_{12}\right)b_2|1\rangle_3\right]c_1|0\rangle_4$$
$$+\left[\left(a_1|00\rangle_{12} + a_2|11\rangle_{12}\right)b_1|1\rangle_3 + \left(a_1|01\rangle_{12} + a_2|10\rangle_{12}\right)b_2|0\rangle_3\right]c_2|1\rangle_4$$

4. $U_5$ on $q_5$ followed by $\text{CNOT}_{5\to 4}$ :
$$\left\{\begin{array}{l}\left[\left(a_1|00\rangle_{12} + a_2|11\rangle_{12}\right)b_1|0\rangle_3 + \left(a_1|01\rangle_{12} + a_2|10\rangle_{12}\right)b_2|1\rangle_3\right]c_1|0\rangle_4 \\ +\left[\left(a_1|00\rangle_{12} + a_2|11\rangle_{12}\right)b_1|1\rangle_3 + \left(a_1|01\rangle_{12} + a_2|10\rangle_{12}\right)b_2|0\rangle_3\right]c_2|1\rangle_4\end{array}\right\}d_1|0\rangle_5$$
$$+\left\{\begin{array}{l}\left[\left(a_1|00\rangle_{12} + a_2|11\rangle_{12}\right)b_1|0\rangle_3 + \left(a_1|01\rangle_{12} + a_2|10\rangle_{12}\right)b_2|1\rangle_3\right]c_1|1\rangle_4 \\ +\left[\left(a_1|00\rangle_{12} + a_2|11\rangle_{12}\right)b_1|1\rangle_3 + \left(a_1|01\rangle_{12} + a_2|10\rangle_{12}\right)b_2|0\rangle_3\right]c_2|0\rangle_4\end{array}\right\}d_2|1\rangle_5$$

$\qquad\qquad\qquad\qquad\qquad\qquad\qquad\qquad\qquad\qquad\qquad\qquad\qquad\qquad\qquad\qquad$ (4)

5. Continue applying $U_k$ on $q_k$ followed by $\text{CNOT}_{k\to k-1}$ iteratively ...
$\qquad \vdots$

n. $U_n$ on $q_n$ followed by $\text{CNOT}_{n\to n-1}$

End of the procedure

where the subscripts on the kets, e.g. $|00\rangle_{12}$, identify the qubits (in the order 1 and 2) represented by the kets. The sequence from $q_1$ to $q_n$ define an order for the $n$ qubits that is associated with the structure in Eq.(4), which will become useful later. In Eq.(4) we have defined the procedure iteratively and it is clear that with only $n-1$ 1-qubit unitaries and $n-1$ CNOT's we have created an n-qubit state with all the qubits entangled. This state is of crucial importance to our later discussion that we will name it the *minimal standard state* $\psi_{\min}^{(n)}$. Although it is difficult to write out the full form of $\psi_{\min}^{(n)}$, the minimal standard state is nonetheless well defined by the iterative procedure in Eq.(4). $\psi_{\min}^{(n)}$ is *minimal* in the sense that among all n-qubit states with all the qubits entangled, $\psi_{\min}^{(n)}$ is the simplest one to create with the described procedure. This is because to add any unentangled qubit to the growing entangled state requires at least one unitary and one CNOT (except the first qubit), which is exactly the procedure described in Eq.(4). $\psi_{\min}^{(n)}$ has two nice properties easily seen from the iterative forms in Eq.(4): 1. the coefficients $C_i^{(k)}$ on each $\phi^{(k)}$ level have only one unique pair, for example $\{a_1, a_2\}$ on the $\phi^{(2)}$ level and $\{c_1, c_2\}$ on the $\phi^{(4)}$ level; 2. the Schmidt decomposition of $\psi_{\min}^{(n)}$ has only $|0\rangle$'s and $|1\rangle$'s on each $\phi^{(k)}$ level, which is much cleaner compared with the general form in Eq.(3). Now to increase the complexity of $\psi_{\min}^{(n)}$ we can either create unique pairs of coefficients $C_i^{(k)}$ on some $\phi^{(k)}$ levels or create terms involving more than $|0\rangle$'s and $|1\rangle$'s in the Schmidt decomposition on each level (e.g. $\phi^{(2)} = a_1 |u_1 v_1\rangle + a_2 |u_2 v_2\rangle$ rather than $a_1 |00\rangle + a_2 |11\rangle$). For the reason that will become obvious after the proof of Theorem 1 below, we only need to focus on the former possibility of creating unique pairs of coefficients. Suppose we want to create a variant pair of $\{a_3, a_4\}$ on the $\phi^{(2)}$ level at the location defined by $|110...0\rangle_{345...n}$ (the subscript in $|\ \rangle_{345...n}$ means the state in the ket corresponds to qubits 3 through $n$ in that order, $0...0$ means all zero throughout these qubits), we can apply $\text{CNOT}_{2\rightarrow1}$ to disentangle the first two qubits, then apply a controlled-unitary operation $C^{n-2}(U)|110...0\rangle_{(345...n)\rightarrow2}$ controlled by qubits 3 through $n$ with the state $|110...0\rangle_{345...n}$, on the target $q_2$ to rotate $a_1 |0\rangle_2 + a_2 |1\rangle_2$ into $a_3 |0\rangle_2 + a_4 |1\rangle_2$, finally apply another $\text{CNOT}_{2\rightarrow1}$ to entangle the first two qubits and we will have the variant pair of $\{a_3, a_4\}$ at the location defined by $|110...0\rangle_{345...n}$. Note any variant pair introduced by this way is normalized in the sense that $|a_3|^2 + |a_4|^2 = 1$. The cost of introducing one variant pair to the $\phi^{(2)}$ level is two CNOT's plus one $C^{n-2}(U)$ gate, which is polynomial in $n$. It is easy to generalize this procedure to introducing a variant pair on any $\phi^{(k)}$ level:

$$\text{CNOT}_{k\rightarrow k-1}, \text{ then } C^{n-k}(U)|\ \rangle_{(k+1...n)\rightarrow k}, \text{ then } \text{CNOT}_{k\rightarrow k-1} \tag{5}$$

where the cost of introducing one variant pair to the $\phi^{(k)}$ level is two CNOT's plus one $C^{n-k}(U)$ gate, which requires fewer gates than $C^{n-2}(U)$ and thus also polynomial in $n$.

**Definition 1.** We define the combined procedure of one application of Eq.(4) and an arbitrary number of applications of Eq.(5) to be the *standard procedure.* We define an arbitrary quantum state that can be created by the standard procedure the *standard state* $\psi^{(n)}$, which is essentially of the form in Eq.(4) but with arbitrary number of variant pairs of coefficients. It is obvious that the simplest standard state $\psi^{(n)}$ is indeed the minimal standard state $\psi_{\min}^{(n)}$.

Here we see that:

**Statement 1.** Any standard state $\psi^{(n)}$ with polynomial number of variant pairs of coefficients compared to the minimal standard state $\psi^{(n)}_{\min}$ can be created in polynomial steps with the standard procedure, and therefore is a quantum state with polynomial creation complexity. We define such states the polynomial standard states $\psi^{(n)}_{poly}$.

$\psi^{(n)}_{poly}$ in Statement 1 defines another subset of quantum states with polynomial creation complexity. Compared to the previously discussed subset of states with polynomial number of non-zero elements, this subset is significantly more general as it includes those quantum states with arbitrary dimension and the maximum number of qubits entangled. So far Statement 1 only applies to any standard state $\psi^{(n)}$, but as we have seen in Eq (3), a general quantum state looks very different from a standard state. Statement 1 can become a useful characterization tool only if it can be generalized to an arbitrary quantum state, and that is achieved by Theorem 1.

**Theorem 1.** Any arbitrary n-qubit quantum state can be transformed into an (n+1)-qubit standard state $\psi^{(n+1)}$ with n steps.

**Proof of Theorem 1.** Take an arbitrary n-qubit quantum state $\phi^{(n)}$, pick any qubit from it, say the first qubit $q_1$ for example, separate the terms associated with $|0\rangle_1$ and $|1\rangle_1$ into two groups:

$$\phi^{(n)} = k_1 \phi_1^{(n-1)} |0\rangle_1 + k_2 \phi_2^{(n-1)} |1\rangle_1 \tag{6}$$

where $\phi_1^{(n-1)}$ and $\phi_2^{(n-1)}$ are (n-1)-qubit states. Expanding the $k_1\phi_1^{(n-1)}$ and $k_2\phi_2^{(n-1)}$ in Eq.(6) into basis states containing all qubits except the first one we have:

$$
\begin{aligned}
\phi^{(n)} =& \left( a_1 \left|000...0\right\rangle_{234...n} + a_2 \left|100...0\right\rangle_{234...n} + a_3 \left|010...0\right\rangle_{234...n} \right. \\
& \left. + a_4 \left|110...0\right\rangle_{234...n} + ... + a_{2^{n-1}} \left|111...1\right\rangle_{234...n} \right) |0\rangle_1 \\
& + \left( b_1 \left|000...0\right\rangle_{234...n} + b_2 \left|100...0\right\rangle_{234...n} + b_3 \left|010...0\right\rangle_{234...n} \right. \\
& \left. + b_4 \left|110...0\right\rangle_{234...n} + ... + b_{2^{n-1}} \left|111...1\right\rangle_{234...n} \right) |1\rangle_1 \\
=& \left( a_1 |0\rangle_1 + b_1 |1\rangle_1 \right)\left|000...0\right\rangle_{234...n} + \left( a_2 |0\rangle_1 + b_2 |1\rangle_1 \right)\left|100...0\right\rangle_{234...n} \\
& + \left( a_3 |0\rangle_1 + b_3 |1\rangle_1 \right)\left|010...0\right\rangle_{234...n} + \left( a_4 |0\rangle_1 + b_4 |1\rangle_1 \right)\left|110...0\right\rangle_{234...n} \\
& + ... + \left( a_{2^{n-1}} |0\rangle_1 + b_{2^{n-1}} |1\rangle_1 \right)\left|111...1\right\rangle_{234...n}
\end{aligned}
\tag{7}
$$

where the order of the basis states is opposite the conventional order of incrementing the last qubit first and then moving to the second last qubit: here we increment $q_2$ first and then move on to $q_3$.

Now if we add one more qubit $q_{n+1}$ to the entanglement by applying a $\text{CNOT}_{1 \to n+1}$ on $\phi^{(n)}$, and group the terms in a leveled manner like Eq.(4), we get:

$$\phi^{(n+1)} = \left\{ \begin{matrix} \left\{ \begin{bmatrix} \left(a_1 \left|00\right\rangle_{n+1,1} + b_1 \left|11\right\rangle_{n+1,1}\right) \left|0\right\rangle_2 + \left(a_2 \left|00\right\rangle_{n+1,1} + b_2 \left|11\right\rangle_{n+1,1}\right) \left|1\right\rangle_2 \end{bmatrix} \left|0\right\rangle_3 \\ + \begin{bmatrix} \left(a_3 \left|00\right\rangle_{n+1,1} + b_3 \left|11\right\rangle_{n+1,1}\right) \left|0\right\rangle_2 + \left(a_4 \left|00\right\rangle_{n+1,1} + b_4 \left|11\right\rangle_{n+1,1}\right) \left|1\right\rangle_2 \end{bmatrix} \left|1\right\rangle_3 \end{matrix} \right\} \left|0\right\rangle_4 \\ + \left\{ \quad\quad\quad\quad\quad\quad\quad \dots\dots \quad\quad\quad\quad\quad\quad\quad \right\} \left|1\right\rangle_4 \end{matrix} \right\} \dots \quad (8)$$

Now apply sequentially $\text{CNOT}_{2\to1}$, $\text{CNOT}_{3\to2}$, …, $\text{CNOT}_{n\to n-1}$ on $\phi^{(n)}$ and we obtain a form similar to Eq.(4) with only differences in the coefficients – after normalizing the coefficients it becomes a standard state $\psi^{(n+1)}$. Note that the steps from Eq.(6) and Eq. (7) are mental steps to group the basis states in a particular way, not actual quantum operations. The actual quantum gates involved are simply n CNOT's. This concludes the proof of Theorem 1. *Theorem 1 is a significant result because it shows an arbitrary quantum state $\phi^{(n)}$, as complex as it may be (such as the form in Eq. (3)), can be transformed to a standard state $\psi^{(n+1)}$ with an almost negligible cost of n CNOT's.* Consequently for any $\phi^{(n)}$, we can check the associated $\psi^{(n+1)}$ against Statement 1 to see if $\phi^{(n)}$ can be created by the standard procedure with polynomial steps. In the next section we discuss an efficient procedure that can check if any standard state belongs to the subset of polynomial standard states.

*3. The procedure to determine if a standard state is polynomial.* Given an arbitrary candidate state, by Theorem 1 we can assume it is in the form of a standard state. Suppose we are allowed to retrieve the coefficient associated with each basis state (e.g. we call the basis state $\left|01010\right\rangle$ for a 5-qubit state, and get the coefficient $C_{01010}$), can we determine if the candidate state belongs to the subset of the polynomial standard states? We know that if the candidate state is a polynomial standard state $\psi_{poly}^{(n)}$, it has most of its coefficients the same as the minimal standard state $\psi_{\min}^{(n)}$, and otherwise it has many of its coefficients different from $\psi_{\min}^{(n)}$. So if we are allowed to retrieve all $2^n$ coefficients, can we just compare them to $\psi_{\min}^{(n)}$? The answer is no! Even if we are allowed to spend exponential ($2^n$) steps to check all the coefficients, the answer is not so simple, because we do not know the values of the coefficients of $\psi_{\min}^{(n)}$, but only know some patterns among the coefficients. As shown in Eq. (4) the coefficients $C_i^{(k)}$ on each $\phi^{(k)}$ level have only one unique pair, for example $\{a_1, a_2\}$ on the $\phi^{(2)}$ level and $\{c_1, c_2\}$ on the $\phi^{(4)}$ level. However to see these simple patterns we need to first put all the $n$ qubits into the correct order to form the correct levels, and in the "wrong" order $\psi_{\min}^{(n)}$ would not show the simple patterns. For example:

$$\psi_{\min}^{(3)} = \left(a_1 |00\rangle_{12} + a_2 |11\rangle_{12}\right) b_1 |0\rangle_3 + \left(a_1 |01\rangle_{12} + a_2 |10\rangle_{12}\right) b_2 |1\rangle_3$$

$$= \left(\frac{a_1 b_1}{A_1} |00\rangle_{13} + \frac{a_2 b_2}{A_1} |11\rangle_{13}\right) A_1 |0\rangle_2 + \left(\frac{a_1 b_2}{A_2} |01\rangle_{13} + \frac{a_2 b_1}{A_2} |10\rangle_{13}\right) A_2 |1\rangle_2 \tag{9}$$

where $A_1 = \sqrt{|a_1 b_1|^2 + |a_2 b_2|^2}$ and $A_2 = \sqrt{|a_1 b_2|^2 + |a_2 b_1|^2}$ are normalization constants. From Eq. (9) we see $\psi_{\min}^{(3)}$ expanded in the order of $|\ \rangle_{123}$ (the first line) shows the pattern of having only one unique pair of coefficients in the innermost level $-\{a_1, a_2\}$ on the $\phi^{(2)}$ level. However, $\psi_{\min}^{(3)}$ expanded in the order of $|\ \rangle_{132}$ (the second line) does not show any obvious pattern among the coefficients. This is potentially problematic because even if the candidate state is the $\psi_{\min}^{(n)}$ itself, without knowing the correct order of the qubits we may need to try all the permutations of n qubits before a simple coefficient pattern can be observed, and a permutation of n qubits introduces an exponential cost of $n! > 2^n$. To solve this problem, we first propose a procedure to detect the coefficient patterns of a $\psi_{\min}^{(n)}$, and later extend it to any $\psi^{(n)}$:

**Procedure 1:**

1. Suppose we are given a $\psi_{\min}^{(n)}$ in the form in Eq. (4) which defines an order of the qubits $q_1$, $q_2$, ..., $q_n$ counting from the innermost level to the outermost level. This order however is unknown to us, so we randomly pick a trio of three qubits $q_j$, $q_k$ and $q_m$ and retrieve the coefficients $C_{00}$, $C_{11}$, $C_{01}$, $C_{10}$ associated with four basis states $|00\rangle_{jk} |0\rangle_m |...\rangle$, $|11\rangle_{jk} |0\rangle_m |...\rangle$, $|01\rangle_{jk} |1\rangle_m |...\rangle$, $|10\rangle_{jk} |1\rangle_m |...\rangle$, where $|...\rangle$ represents a basis state of the qubits other than the selected three. $|...\rangle$ can be randomly chosen from all $2^{n-4}$ possible states but has to be the same for all four basis states listed.

2. Now if we happen to hit $q_1$ with $q_j$, $q_2$ with $q_k$, and $q_3$ with $q_m$ ($j=1$, $k=2$, $m=3$), then we have $C_{00} = a_1 b_1 h$, $C_{11} = a_2 b_1 h$, $C_{01} = a_1 b_2 h$, $C_{11} = a_2 b_2 h$, where $a_1$, $b_1$, $a_2$, $b_2$ are taken from the first two levels, $h = cd...$ is the collective coefficient of the remaining levels. Since $h$ is the same for all four coefficients, we have a ratio pattern $\dfrac{C_{00}}{C_{11}} = \dfrac{C_{01}}{C_{10}}$.

3. Relaxing the condition in Step 2 to $q_k$ being in the middle of $q_j$ and $q_m$ ($j < k < m$ or $m < k < j$), we will still have the ratio pattern $\dfrac{C_{00}}{C_{11}} = \dfrac{C_{01}}{C_{10}}$. This is because for example, if $j < k < m$, for $|00\rangle_{jk} |0\rangle_m |...\rangle$ and $|11\rangle_{jk} |0\rangle_m |...\rangle$, all the qubits in the following ranges are the same: from the last qubit (include) to $q_k$ (exclude), from $q_k$ (exclude) to $q_j$ (exclude), and from $q_j$ (exclude) to the first qubit (include). Consequently, $\frac{C_{00}}{C_{11}}$ cancels out all the coefficients associated with the qubits that are the same, and only shows the difference in the coefficients due to $|00\rangle_{jk}$ and $|11\rangle_{jk}$ being different. Similarly for $|01\rangle_{jk}|1\rangle_m|...\rangle$ and $|10\rangle_{jk}|1\rangle_m|...\rangle$, $\frac{C_{01}}{C_{10}}$ cancels out all the coefficients associated with the qubits that are the same, and only shows the difference in the coefficients due to $|01\rangle_{jk}$ and $|10\rangle_{jk}$ being different. Now it is easy to show that $\frac{C_{00}}{C_{11}} = \frac{C_{01}}{C_{10}}$ holds. The same argument also applies to the case where $m < k < j$.

4. With the same argument in Step 3 we can obtain the ratio pattern for $j < m < k$ and $k < m < j$ is $\frac{C_{00}}{C_{10}} = \frac{C_{01}}{C_{11}}$ and the ratio pattern for $m < j < k$ and $k < j < m$ is $\frac{C_{00}}{C_{11}} = \frac{C_{10}}{C_{01}}$. A concrete example of the ratio pattern tests is shown in the Supplementary Information.

5. Combining the results of Steps 3 and 4, for any random pick of a trio of three qubits $q_j$, $q_k$ and $q_m$ we can retrieve and check the four coefficients $C_{00}$, $C_{11}$, $C_{01}$, $C_{10}$ to determine which of three ratio patterns is satisfied. Then by the satisfied ratio pattern we will know which of $q_j$, $q_k$ and $q_m$ is in the middle according to the unknown order defined by $\psi_{\min}^{(n)}$ in the form in Eq. (4). We can repeat this process for different choices of trios $q_j$, $q_k$ and $q_m$ and "sequence" more qubits until we discover the unknown order of all the qubits. For example say in the first trial we discover that $j < m < k$ (or $k < m < j$), we can then determine the ratio pattern for a new trio with $q_j$, $q_k$, plus a new qubit $q_l$ in the second trial. If $q_k$ is in the middle then we conclude the sequence as $j < m < k < l$ (or $l < k < m < j$); if $q_j$ is in the middle then we conclude the sequence as $l < j < m < k <$ (or $k < m < j < l$). If $q_l$ is in the middle then the sequence could be either $j < l < m < k$ (or $k < m < l < j$) or $j < m < l < k$ (or $k < l < m < j$), which can then be decided by running a third trial with the trio of $q_j$, $q_l$ and $q_m$. Repeating this process allows us to sequence all the qubits into two possible orders with one order being the exact reverse of the other. It can be shown (see the Supplementary Information) that knowing a qubit sequence of length $h$, we can find the correct position of a new qubit with a maximum of $floor\left(\frac{h+1}{2}\right)$ trials, and therefore the total number of trials required to sequence all $n$ qubits is polynomial in $n$: $O\left(n^2\right)$.

6. Once the correct order of the qubits for $\psi_{\min}^{(n)}$ in the form in Eq. (4) is known, we effectively know the structure of $\psi_{\min}^{(n)}$, and the unique coefficient pair on each level of Eq. (4) can be obtained from two coefficient retrievals and one division (see the Supplementary Information).

*Procedure 1 is significant because for any given $\psi_{\min}^{(n)}$ it allows us to determine its structure in the form of Eq. (4) along with all the coefficients on all levels such that we can construct the minimal method to create $\psi_{\min}^{(n)}$. It is remarkable that Procedure 1 carries only a polynomial cost as it only retrieves selective coefficients from the state and does not scan through all the coefficients.* As shown next this procedure will also allow us to determine if a standard state is polynomial.

Suppose now we are given a state in the form of a standard state $\psi^{(n)}$, if it is a polynomial standard state then it contains at most $cn^k \ll 2^n$ number of variant coefficients as compared to the $\psi_{\min}^{(n)}$. Consequently, if we carry out Procedure 1 on $\psi^{(n)}$ we should have an extremely high probability

$$p_0 = \left(1 - \frac{cn^k}{2^n}\right)^{N_0} \approx 1$$ to notice no difference from a $\psi_{\min}^{(n)}$ such that all trials involved will succeed in finding a ratio pattern among the coefficients, where $N_0 = O(n^2)$ is the number of trials involved in Procedure 1. The contrapositive is also true that, if we do encounter one failure to get a ratio pattern within $N_0$ trials, then $\psi^{(n)}$ is extremely unlikely to be a polynomial standard state. To make this observation strict we may use the method of Bayesian inference. Suppose the probability of failing to get any of the ratio pattern from Procedure 1 in one trial is $p$, then the probability of the event of the first failure happening at the $N^{\text{th}}$ trial is $P(x \mid p) = p(1-p)^{N-1}$, where $x$ represents the event. Now we want to infer the posterior probability $P(p \le p_1 \mid x)$, i.e. the probability of $p \le p_1$ given the event $x$ has happened. According to Bayesian inference this can be done by integrating the posterior distribution function over the interval of 0 to $p_1$. The posterior distribution is:

$$f(p \mid x) = \frac{P(x \mid p) f(p)}{\int_0^1 P(x \mid p) f(p) \, dp} = \frac{p(1-p)^{N-1}}{\int_0^1 \left[ p(1-p)^{N-1} \right] dp} \tag{10}$$

where the prior distribution function is the uniform one $f(p) = 1$ (uniform because we assume no prior knowledge of $p$). Integrating $f(p \mid x)$ from 0 to $p_1$ gives:

$$P(p \le p_1 \mid x) = \int_0^{p_1} f(p \mid x) \, dp = 1 - (Np_1 + 1)(1 - p_1)^N \tag{11}$$

The probability $p$ is directly related to $K$ – the number of variant coefficients in $\psi^{(n)}$ as compared to $\psi_{\min}^{(n)}$ – such that $p \approx \frac{K}{2^n}$ (Depending on the location of the variants relative to the structure of $\psi_{\min}^{(n)}$, the exact number may vary by a small factor $\alpha$, but it is unimportant because $\alpha$ is negligible compared to both $K$ and $2^n$), and the event of $K \leq K_1$ is equivalent to the event of $p \leq p_1 = \dfrac{K_1}{2^n}$. By Eq. (11) we see that if the first failure to get a ratio pattern happens on the $N^{\text{th}}$ trial, the probability of $p \leq p_1 = \dfrac{K_1}{2^n}$ is $1 - \left( \dfrac{NK_1}{2^n} + 1 \right)\left( 1 - \dfrac{K_1}{2^n} \right)^N$, which is close to zero when $N \leq N_0$ and $K_1 \leq cn^k \ll 2^n$. This formally establishes the fact that if we encounter the first failure to get a ratio pattern within $N_0$ trials, we can terminate the process and conclude $\psi^{(n)}$ is extremely unlikely to be a polynomial standard state as defined by having at most $cn^k \ll 2^n$ number of variant coefficients compared to the $\psi_{\min}^{(n)}$. On the other hand if we keep doing the trials and encounter no failure of getting a ratio pattern after some very large number $N$ of trials, we can then infer the posterior probability $P\left( p \leq p_1 \mid y \right)$, where $y$ is the event of always getting a ratio pattern over $N$ trials. Again the posterior distribution is:

$$f\left( p \mid y \right) = \frac{P\left( y \mid p \right) f\left( p \right)}{\int_0^1 P\left( y \mid p \right) f\left( p \right) dp} = \frac{\left( 1 - p \right)^N}{\int_0^1 \left( 1 - p \right)^N dp} \tag{12}$$

Integrating $f\left( p \mid y \right)$ from 0 to $p_1$ gives:

$$P\left( p \leq p_1 \mid y \right) = \int_0^{p_1} f\left( p \mid y \right) dp = 1 - \left( 1 - p_1 \right)^{N+1}$$
$$P\left( K \leq K_1 \mid y \right) = 1 - \left( 1 - \frac{K_1}{2^n} \right)^{N+1} \tag{13}$$

Eq. (13) means that even if $K_1 \ll 2^n$, with a very large number $N$ of trials all being successful, the probability of $K \leq K_1$ can still be an appreciable number. For example, if we have 50 qubits ($n = 50$), "polynomial" is defined as $K \leq K_1 = 2^{30}$, for a given $\psi^{(n)}$ we have $N = 10^7$ trials all being successful, then we can conclude with $1 - 10^{-4}$ confidence probability that $\psi^{(n)}$ is polynomial. Although $N = 10^7$ is a very large number, it is still much smaller than $2^{30} \approx 10^9$, and thus can be considered as "polynomial". This however will not work if we define "polynomial" with a smaller cutoff number $K_1$ because the number of trials needed to reach a high confidence probability may be far greater than $K_1$. Nonetheless we note that $\dfrac{2^{30}}{2^{50}} < 10^{-6}$, so a state $\psi^{(n)}$ with $K < 2^{30}$ variant coefficients from the $\psi_{\min}^{(n)}$ is almost equal to $\psi_{\min}^{(n)}$ in the sense that $\left\| \psi^{(n)} - \psi_{\min}^{(n)} \right\| < 10^{-6}$. Since if $K < 2^{30}$ we have extremely high probability $p_0 \geq \left( 1 - \dfrac{2^{30}}{2^{50}} \right)^{N_0} \approx 1$ to

complete all $N_0$ steps in Procedure 1, we can discover the structure and all the coefficients of $\psi_{\min}^{(n)}$ and therefore can create $\psi_{\min}^{(n)}$ with the minimal procedure described in Eq. (4). The $\psi_{\min}^{(n)}$ created is then a good approximation to the original $\psi^{(n)}$.

*4. The procedure to determine the variant coefficients.* The previous section has presented a procedure to determine if a candidate $\psi^{(n)}$ is a polynomial standard state defined by some $\psi_{\min}^{(n)}$ using selective coefficient retrieval and Bayesian inference. However the method is only efficient if the cutoff number $K_1$ is sufficiently large. In addition, in some situations it may be desirable to determine the locations and absolute values of the variant coefficients such that $\psi^{(n)}$ can be more accurately reproduced. In the following we show that this is possible with an additional procedure when the quantum state $\psi^{(n)}$ is available in large number of copies.

**Procedure 2:**

1.  Suppose given a $\psi^{(n)}$, we are allowed to retrieve the coefficients associated with specific basis states. In addition we also have the quantum state of $\psi^{(n)}$ in large number of copies. How could we have copies of $\psi^{(n)}$ before its exact form is known? The copies could be created by repeating an experiment many times, or by applying a long sequence of quantum gates that we hope to improve with our new method. This does not violate the no-cloning theorem because the state is indeed known. Now assume we have completed all the steps in Procedure 1 without any difference from a $\psi_{\min}^{(n)}$. This means we have discovered the structure and coefficients on all levels of this $\psi_{\min}^{(n)}$, so the qubits $q_1$, $q_2$, …, $q_n$ are correctly ordered. Now label the levels of $\psi_{\min}^{(n)}$ in the form of Eq. (4) from the innermost level $L_1$, $L_2$, …, to the outermost level $L_{n-1}$ such that $L_1$ is the level with the coefficients $\{a_1, a_2\}$, $L_2$ is the level with the coefficients $\{b_1, b_2\}$, …, and so on.

2.  For the moment we assume that all the possible variant coefficients are on $L_1$. We take the copies of $\psi^{(n)}$ and apply projection measurement on $q_1$ to determine the probability of getting $|0\rangle$ and $|1\rangle$ for the first qubit. If there is no variant coefficient on $L_1$, the result should be the same as $\psi_{\min}^{(n)}$: $|a_1|^2$ for $|0\rangle$ and $|a_2|^2$ for $|1\rangle$, or simply $\left(|a_1|^2, |a_2|^2\right)$. If indeed there are variant coefficients on $L_1$, even if only one pair, the probability will deviate from $\left(|a_1|^2, |a_2|^2\right)$, and should be detected with enough number of projection measurements. In the latter case we continue to Step 3.

3.  We continue the projection measurement on both $q_1$ and $q_2$ to determine the probability of getting $|00\rangle$ and $|01\rangle$ for the first two qubits. The ideal case of $\psi_{\min}^{(n)}$ would give the result of $\left(\left|a_1 b_1\right|^2, \left|a_1 b_2\right|^2\right)$, but as we have already detected variants in Step 2, at least one of the two probabilities should be different from the $\psi_{\min}^{(n)}$ value. If the probability for getting $\left|00\right\rangle$ is the same as $\left|a_1 b_1\right|^2$, there is no variant in the group of basis states containing $\left|00\right\rangle$ for the first two qubits and we do not continue with this group; otherwise if the probability of getting $\left|00\right\rangle$ is not $\left|a_1 b_1\right|^2$, there are variants in the group and we continue with this group. We determine if to continue with the group containing $\left|01\right\rangle$ in the first two qubits by the same method.

4.  If we have decided to continue with the $\left|00\right\rangle$ group in Step 3, now apply projection measurement on $q_1$, $q_2$ and $q_3$ to determine the probability of getting $\left|000\right\rangle$ and $\left|001\right\rangle$. Again the ideal case of $\psi_{\min}^{(n)}$ would give the result of $\left(\left|a_1 b_1 c_1\right|^2, \left|a_1 b_1 c_2\right|^2\right)$, and any difference in the measurement results would indicate the presence of variants in the respective groups containing $\left|000\right\rangle$ and $\left|001\right\rangle$ for the first three qubits. We do the same with the $\left|01\right\rangle$ group if decided to continue with it in Step 3.

5.  We continue the processes described in Steps 2 through 4 for progressively more qubits, until we reach the second-to-last qubit $q_{n-1}$ (due to the unique structure of the standard state, the last qubit $q_n$ is automatically fixed if all the preceding qubits are fixed). The whole process can be represented by a binary tree for which the root node represents the measurement on $\left|0\right\rangle$ for the first qubit, the nodes on the second level represent the measurements on $\left|00\right\rangle$ and $\left|01\right\rangle$, the nodes on the third level represent the measurements on possibly $\left|000\right\rangle$, $\left|001\right\rangle$, $\left|010\right\rangle$ and $\left|011\right\rangle$, and so on. We note that the majority of the nodes on the binary tree are missing because we only continue to expand a node if the group represented by that node has a different probability value from $\psi_{\min}^{(n)}$. Indeed, if the original candidate state $\psi^{(n)}$ has $K$ pairs of variant coefficients compared to $\psi_{\min}^{(n)}$, then the corresponding tree has maximally $K$ branches that reach the last level of $n-1$. Now to give an upper limit to the total number of nodes on the tree, we note that each of the $K$ branches contains $n-1$ nodes on the main path and up to $n-2$ terminating nodes (i.e. leaves). The total number of nodes on the three is therefore $N \le K \cdot (2n-3)$. Note that in practice $N$ is always much smaller than this upper limit because two branches may share many nodes, thus $K \cdot (2n-3)$ is almost always overcounting. Nonetheless $K \cdot (2n-3)$ is polynomial in $n$, which means we can efficiently determine all the variant coefficients if they are on $L_1$ only.

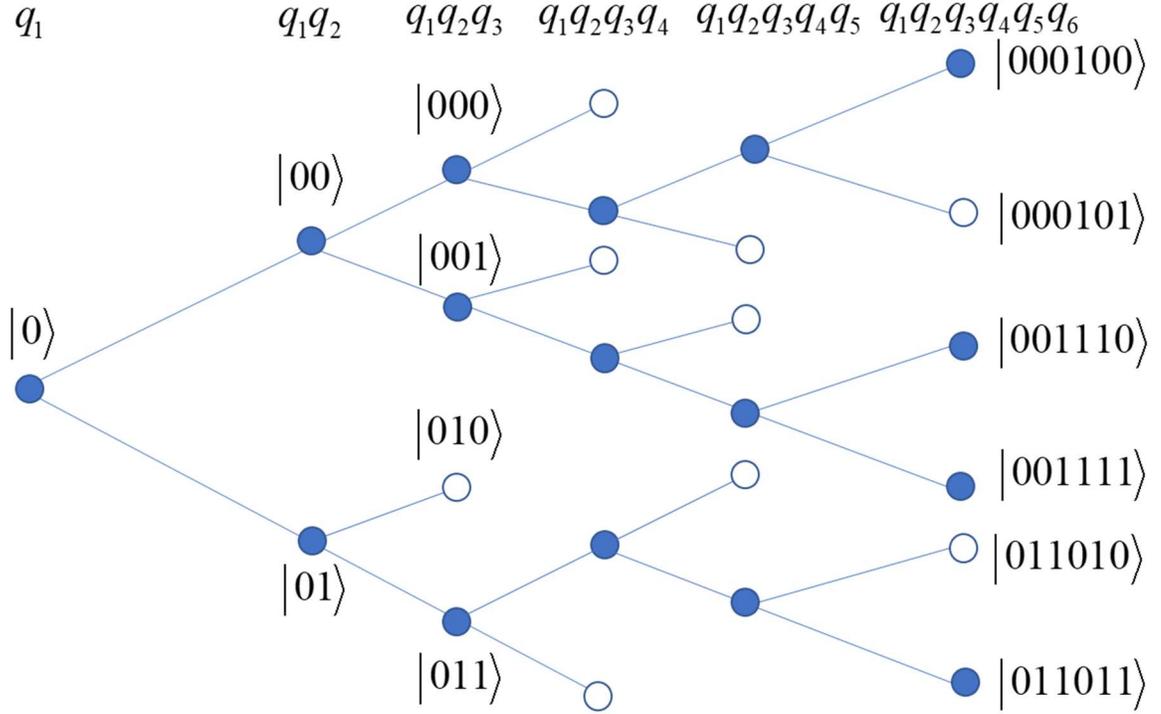

Figure 1. An example of the six-leveled binary tree defined by applying Procedure 2 on a seven-qubit standard state $\psi^{(7)}$. Each node on the tree corresponds to a projection measurement on the specified basis state. At the blue nodes variants are detected and the branches can continue; at the white nodes variants are not detected and the branches terminate. Eventually we reach the end level with three main branches. Four variants are present as indicated by the four blue nodes at the end level.

Figure 1 shows an example of a six-leveled binary tree defined by a series of projection measurements on a seven-qubit $\psi^{(7)}$. We start at the root node of $|0\rangle$ and progress to the right. At the blue nodes variants are detected and the branches can continue; at the white nodes variants are not detected and the branches terminate. Eventually there are three main branches that reach the end level and 25 nodes in total, representing 25 projection measurements. This number is much smaller than $K \cdot (2n-3) = 44$ with $n = 7$ and $K = 4$, because the two variants on $|001110\rangle$ and $|001111\rangle$ share the same branch. This demonstrates that we can determine the locations and values of the variants efficiently with projection measurements.

6. As details shown in the Supplementary Information, the procedure described above can be extended to variant coefficients on all other levels, given that a variant on $L_k$ can only be detected by the measurement involving the qubit $q_k$. Consequently if we have no variant before $L_k$, we will not detect any variants on $L_k$ and higher levels until we expand the measurement to include $q_k$. This requires us to modify the procedure to continue even if no variant is detected in any of the nodes in an entire level. The upper limit of the total number of nodes is still $N \leq K \cdot (2n-3)$ with even more overcounting because typically higher level variants will share a branch with lower level variants.

Procedure 2 allows us to efficiently determine the locations and absolute values (phases cannot be determined) of the variant coefficients on $\psi^{(n)}$ as compared to $\psi_{\min}^{(n)}$ given that a large number of copies of $\psi^{(n)}$ is available. This is possible because projection measurements on a quantum state can automatically sum over the probabilities associated with a large number of coefficients, thus the very few variant coefficients on $\psi^{(n)}$ can be found efficiently given enough measurement precision. The projection measurement used at each node is inherently probabilistic but the standard error of the mean $\sigma_m$ is given by $\sigma_m = \dfrac{\sigma}{\sqrt{\mathcal{N}}}$, where $\sigma$ is the inherent standard deviation defined by the coefficients of $\psi_{\min}^{(n)}$, and $\mathcal{N}$ is the number of data points collected at each node. We see that if $\mathcal{N}$ is large enough we can make $\sigma_m$ much smaller than the difference between $\psi^{(n)}$ and $\psi_{\min}^{(n)}$. In practice, we can fix $\mathcal{N}$ at an acceptable number and only detect those differences between $\psi^{(n)}$ and $\psi_{\min}^{(n)}$ greater than the pre-determined $\sigma_m$, and any difference smaller than $\sigma_m$ is considered an acceptable error for the creation of $\psi^{(n)}$.

## III. Discussion and Conclusion

In this work we have examined the problem of determining the creation complexity of a given candidate quantum state. By using the elementary gate set of all 2-qubit controlled-unitary gates, we relate the polynomial states to the space represented by tuples of $C(U_i)$ with polynomial lengths. The problem of determining if a candidate quantum state is polynomial then becomes determining if a vector in an $\mathbf{R}^{2^{n+1}-2}$ space belongs to the subspace of $\mathbf{R}^{4cn^k}$, which carries an exponential cost. We then show it is possible to define a proper subset of the polynomial states such that simple characteristics can be identified among the coefficients of the states and partial knowledge of the creation complexity can be obtained. The definition of the subset is based on the minimal standard state $\psi_{\min}^{(n)}$ and the standard procedure as defined by combining Eq. (4) and (5). Remarkably, Theorem 1 proves that an arbitrary quantum state $\phi^{(n)}$, as complex as it may be (such as the form in Eq. (3)), can be transformed to a standard state $\psi^{(n+1)}$ with an almost negligible cost of n CNOT's. This result is essential as it allows us to transform any arbitrary $\phi^{(n)}$ into the associated $\psi^{(n+1)}$ and then determine if $\phi^{(n)}$ belongs to the subset of polynomial states defined by $\psi_{\min}^{(n+1)}$. This leads to a method where we efficiently identify the presence (or absence) of the ratio patterns among the coefficients of $\psi^{(n)}$ by the combination of the coefficient sampling procedure of Procedure 1 and Bayesian inference. Note that when the method finds $\psi^{(n)}$ to belong to the subset, it also automatically generates a minimal procedure to create the corresponding $\psi^{(n)}_{\min}$ as a good approximation of $\psi^{(n)}$. Finally we demonstrate a method to determine the locations and absolute values of the variant coefficients of $\psi^{(n)}$ compared to $\psi^{(n)}_{\min}$ using projection measurements. The subset of polynomial states based on $\psi^{(n)}_{\min}$ is general in the sense that it contains the states with arbitrary dimension and the maximum entanglement allowed by the number of qubits. Consequently the subset should include a large number of diverse choices of quantum states that may be useful in quantum computing operations. The partial knowledge of a state's creation complexity as obtained by our methods can therefore be used to guide the development of quantum algorithms as it greatly reduces the search space. In addition, the idea of identifying common features among the coefficients of a candidate state and sampling the coefficients for these features may be further developed in a future study for more complete characterization of the quantum states with polynomial creation complexity, further improving design of quantum circuits and algorithms.

## IV.    Acknowledgement

Z. H. would like to thank Peng Zhou for insightful discussions. Z. H. and S. K. would like to acknowledge funding by the U.S. Department of Energy (Office of Basic Energy Sciences) under Award No. DE-SC0019215.

## V.    Supplementary information

Supplementary information is available after the reference list.

# Supplementary information: Characterization of quantum states based on creation complexity

Zixuan Hu and Sabre Kais*

*Department of Chemistry, Department of Physics, and Birck Nanotechnology Center, Purdue University, West Lafayette, IN 47907, United States*
*Email:* kais@purdue.edu

This supplementary document supports the discussion in the main text by providing technical details. Section 1 provides an example for Procedure 1 in the main text. Section 2 provides an example for Procedure 2 in the main text to illustrate how to determine variant coefficients on higher levels.

## 1. An example for Procedure 1 in the main text.

In the main text we have presented Procedure 1 for determining the coefficient pattern of a minimal standard state $\psi_{\min}^{(n)}$. The discussion there is a general one involving abstract arguments. Here we provide a concrete example for a five-qubit $\psi_{\min}^{(5)}$. Suppose we are given a state known to be a minimal standard state of five qubits:

$$\psi_{\min}^{(5)} = \begin{cases} \left\{ \begin{array}{l} \left[ \left( a_1 \left| 00 \right\rangle_{12} + a_2 \left| 11 \right\rangle_{12} \right) b_1 \left| 0 \right\rangle_3 + \left( a_1 \left| 01 \right\rangle_{12} + a_2 \left| 10 \right\rangle_{12} \right) b_2 \left| 1 \right\rangle_3 \right] c_1 \left| 0 \right\rangle_4 \\ + \left[ \left( a_1 \left| 00 \right\rangle_{12} + a_2 \left| 11 \right\rangle_{12} \right) b_1 \left| 1 \right\rangle_3 + \left( a_1 \left| 01 \right\rangle_{12} + a_2 \left| 10 \right\rangle_{12} \right) b_2 \left| 0 \right\rangle_3 \right] c_2 \left| 1 \right\rangle_4 \end{array} \right\} d_1 \left| 0 \right\rangle_5 \\ + \left\{ \begin{array}{l} \left[ \left( a_1 \left| 00 \right\rangle_{12} + a_2 \left| 11 \right\rangle_{12} \right) b_1 \left| 0 \right\rangle_3 + \left( a_1 \left| 01 \right\rangle_{12} + a_2 \left| 10 \right\rangle_{12} \right) b_2 \left| 1 \right\rangle_3 \right] c_1 \left| 1 \right\rangle_4 \\ + \left[ \left( a_1 \left| 00 \right\rangle_{12} + a_2 \left| 11 \right\rangle_{12} \right) b_1 \left| 1 \right\rangle_3 + \left( a_1 \left| 01 \right\rangle_{12} + a_2 \left| 10 \right\rangle_{12} \right) b_2 \left| 0 \right\rangle_3 \right] c_2 \left| 0 \right\rangle_4 \end{array} \right\} d_2 \left| 1 \right\rangle_5 \end{cases} \quad (14)$$

Suppose the structure in Eq. (14) is unknown to us, but we are allowed to retrieve the coefficient associated with any given basis state – e.g. if we call the basis state $\left| 00110 \right\rangle_{12345}$ we get $C_{\left| 00110 \right\rangle} = a_1 b_1 c_2 d_1$. Following Step 1 in the main text, we pick a trio of three ordered qubits $(q_4, q_3, q_1)$, and retrieve the coefficients $C_{00}$, $C_{11}$, $C_{01}$, $C_{10}$ associated with four basis states $\left| 00 \right\rangle_{43} \left| 0 \right\rangle_1 \left| ... \right\rangle_{25}$, $\left| 11 \right\rangle_{43} \left| 0 \right\rangle_1 \left| ... \right\rangle_{25}$, $\left| 01 \right\rangle_{43} \left| 1 \right\rangle_1 \left| ... \right\rangle_{25}$, $\left| 10 \right\rangle_{43} \left| 1 \right\rangle_1 \left| ... \right\rangle_{25}$, where $\left| ... \right\rangle_{25}$ is any basis state that represents the qubits other than the selected three. Say we randomly pick $\left| ... \right\rangle_{25} = \left| 00 \right\rangle_{25}$, then $C_{00} = a_1 b_1 c_1 d_1$, $C_{11} = a_1 b_1 c_2 d_1$, $C_{01} = a_2 b_2 c_1 d_1$, $C_{10} = a_2 b_2 c_2 d_1$. Obviously $\frac{C_{00}}{C_{11}} = \frac{a_1 b_1 c_1 d_1}{a_1 b_1 c_2 d_1} = \frac{C_{01}}{C_{10}} = \frac{a_2 b_2 c_1 d_1}{a_2 b_2 c_2 d_1} = \frac{c_1}{c_2}$, so by the rule described in Step 3 of Procedure 1 in the main text, we conclude $q_3$ is in the middle of $q_4$ and $q_1$ for the order defined by Eq. (14), which is indeed correct. Next we consider the trio of $(q_4, q_1, q_2)$, and and retrieve the coefficients $C_{00}$, $C_{11}$, $C_{01}$, $C_{10}$ associated with four basis states $|00\rangle_{41}|0\rangle_2|...\rangle_{35}$, $|11\rangle_{41}|0\rangle_2|...\rangle_{35}$, $|01\rangle_{41}|1\rangle_2|...\rangle_{35}$, $|10\rangle_{41}|1\rangle_2|...\rangle_{35}$, where again $|...\rangle_{35}$ can be randomly picked to be $|00\rangle_{35}$. We find that $C_{00} = a_1 b_1 c_1 d_1$, $C_{11} = a_2 b_2 c_2 d_1$, $C_{01} = a_2 b_1 c_1 d_1$, $C_{10} = a_1 b_2 c_2 d_1$, and this time the ratio pattern is $\dfrac{C_{00}}{C_{10}} = \dfrac{a_1 b_1 c_1 d_1}{a_1 b_2 c_2 d_1} = \dfrac{C_{01}}{C_{11}} = \dfrac{a_2 b_1 c_1 d_1}{a_2 b_2 c_2 d_1} = \dfrac{b_1 c_1}{b_2 c_2}$. By Step 4 of Procedure 1, we conclude $q_2$ is in the middle of $q_4$ and $q_1$ which is again correct. We repeat the process for the trio $(q_4, q_1, q_5)$, and get $\dfrac{C_{00}}{C_{11}} = \dfrac{a_1 b_1 c_1 d_1}{a_2 b_2 c_2 d_1} = \dfrac{C_{10}}{C_{01}} = \dfrac{a_1 b_1 c_1 d_2}{a_2 b_2 c_2 d_2} = \dfrac{a_1 b_1 c_1}{a_2 b_2 c_2}$, which correctly implies $q_4$ is in the middle of $q_5$ and $q_1$. Now with another trial involving the trio of $(q_1, q_3, q_2)$, we conclude $q_2$ is in the middle of $q_3$ and $q_1$. So after all these trials we find the correct order of the five qubits is $(q_1, q_2, q_3, q_4, q_5)$ (or $(q_5, q_4, q_3, q_2, q_1)$, but it does not matter). Next we retrieve the coefficients associated with $|00000\rangle_{12345}$ and $|11000\rangle_{12345}$: $C_{|00000\rangle} = a_1 b_1 c_1 d_1$ and $C_{|11000\rangle} = a_2 b_1 c_1 d_1$, and taking the ratio $\dfrac{C_{|00000\rangle}}{C_{|11000\rangle}} = \dfrac{a_1}{a_2}$ will allow us to calculate the coefficient pair of the first level $a_1$ and $a_2$. Similarly the coefficient pairs on all other levels can be found by retrieving two coefficients and taking the ratio. This apparent simplicity is guaranteed by knowing the correct order of the five qubits, and we see that Procedure 1 is very useful as it allows us to efficiently determine both the structure and the creation procedure of a $\psi_{\min}^{(n)}$.

We have just provided a concrete example of Procedure 1 in the main text to show the power of it to sequence the five qubits of a $\psi_{\min}^{(5)}$ into the correct order thus allowing coefficient determination. In the main text we claim that for an n-qubit $\psi_{\min}^{(n)}$ the total number of trials required to sequence all $n$ qubits is polynomial in $n$: $O(n^2)$. Here we provide a concrete explanation with the 5-qubit example. Suppose with the first trio we sequence 3 qubits in the correct order $(q_1, q_2, q_4)$, now given any new qubit we want to insert it into the sequence with the correct order. For example if the new qubit is $q_3$ we want to insert it between $q_2$ and $q_4$: $(q_1, q_2, q_3, q_4)$; if the new qubit is $q_5$ we want to insert it to the right of $q_4$: $(q_1, q_2, q_4, q_5)$. Suppose the new qubit is $q_5$, we choose the trio involving the first qubit of the known sequence, the last qubit of the known sequence, and the new qubit: $(q_1, q_4, q_5)$. As described above this trio will inform us that $q_5$ is to the right of $q_4$. As there is currently no other qubit to the right of $q_4$, $q_5$ can be unambiguously inserted to the sequence as $(q_1, q_2, q_4, q_5)$. The next new qubit is $q_3$, we again first choose the trio involving the first qubit of the known sequence, the last qubit of the known sequence, and the new qubit:

$(q_1, q_5, q_3)$. This will inform us that $q_3$ is in the middle of $q_1$ and $q_5$. As there are two other qubits $q_2$ and $q_4$ in the middle of $q_1$ and $q_5$, the correct position of $q_3$ is not yet decided. Next we choose the trio involving the second qubit of the known sequence, the second last qubit of the known sequence, and the new qubit: $(q_2, q_4, q_3)$. This will inform us that $q_3$ is in the middle of $q_2$ and $q_4$. As there is no additional qubit in the middle of $q_2$ and $q_4$, we can conclude the 5-qubit sequence $(q_1, q_2, q_3, q_4, q_5)$ is in the correct order. From this example we can see that in general, given a known sequence of $h$ qubits, we can insert a new qubit into the sequence by testing trios involving first (the first qubit, the last qubit, the new qubit), then (the second qubit, the second last qubit, the new qubit), then (the third qubit, the third last qubit, the new qubit), ... and so on. If at any step the new qubit is not in the middle of the other two qubits, we find its correct insertion position and conclude the trial with a sequence of $h+1$ qubits. If indeed the new qubit is always in the middle of the two other qubits, then at the maximum we will need $floor\left(\dfrac{h+1}{2}\right)$ trials to insert the new qubit into the correct place and proceed with a sequence of $h+1$ qubits. For a general n-qubit state, we start with a known sequence of three qubits, and the maximum total number of trials needed to completely sequence all n qubits is $\displaystyle\sum_{h=3}^{n} floor\left(\dfrac{h+1}{2}\right) = O\left(n^2\right)$.

## 2. An example for Procedure 2 in the main text.

In the main text we have presented Procedure 2 for determining the locations and values of the variant coefficients of a $\psi^{(n)}$ on the first level $L_1$. Here we show how variant coefficients on higher levels can be detected. Suppose we are given a 5-qubit standard state $\psi^{(5)}$ that has only one variant coefficient pair on the third level $L_3$ as compared to the $\psi^{(5)}_{\min}$ in Eq. (14):

$$\psi^{(5)} = \begin{pmatrix} \left\{ \begin{aligned} &\left[\left(a_1|00\rangle_{12} + a_2|11\rangle_{12}\right)b_1|0\rangle_3 + \left(a_1|01\rangle_{12} + a_2|10\rangle_{12}\right)b_2|1\rangle_3\right]c_3|0\rangle_4 \\ &+\left[\left(a_1|00\rangle_{12} + a_2|11\rangle_{12}\right)b_1|1\rangle_3 + \left(a_1|01\rangle_{12} + a_2|10\rangle_{12}\right)b_2|0\rangle_3\right]c_4|1\rangle_4 \end{aligned} \right\}d_1|0\rangle_5 \\ +\left\{ \begin{aligned} &\left[\left(a_1|00\rangle_{12} + a_2|11\rangle_{12}\right)b_1|0\rangle_3 + \left(a_1|01\rangle_{12} + a_2|10\rangle_{12}\right)b_2|1\rangle_3\right]c_1|1\rangle_4 \\ &+\left[\left(a_1|00\rangle_{12} + a_2|11\rangle_{12}\right)b_1|1\rangle_3 + \left(a_1|01\rangle_{12} + a_2|10\rangle_{12}\right)b_2|0\rangle_3\right]c_2|0\rangle_4 \end{aligned} \right\}d_2|1\rangle_5 \end{pmatrix} \quad (15)$$

where the variant pair is $\{c_3, c_4\}$. Here we assume $|c_3|^2 + |c_4|^2 = 1$ because the way the standard state is defined in the main text guarantees all the coefficient pairs on each level are normalized. From Procedure 1 we have determined the structure of $\psi^{(5)}_{\min}$, but we do not know the existence, the location, and the values of $\{c_3, c_4\}$. Now following Step 2 of Procedure 2 we first take the copies of $\psi^{(n)}$ and apply projection measurement on $q_1$ to determine the probability of getting $|0\rangle$ and $|1\rangle$ for the first qubit. As there is no variant coefficient on $L_1$, this measurement will give the same result as $\psi_{\min}^{(5)}$: $|a_1|^2$ for $|0\rangle$ and $|a_2|^2$ for $|1\rangle$, or simply $\left(|a_1|^2, |a_2|^2\right)$. The reason why this measurement cannot detect the variant coefficients on $L_3$ is that the projection measurement on $q_1$ alone can only distinguish the probabilities given by the coefficients associated with $q_1$, and the probabilities given by the coefficients associated with other qubits are automatically summed over. Consequently, through simple derivation we find that a measurement involving $q_1$ through $q_k$ can only detect variant coefficients on $L_1$ through $L_k$, but not $L_{k+1}$ and beyond. Now back to our example, the first step of measuring $q_1$ alone does not find any variants, and we continue with Step 3 of Procedure 2 by measuring the probability of getting $|00\rangle$ and $|01\rangle$ for the first two qubits. As the variant pair is on $L_3$, this step will also give the same result as $\psi_{\min}^{(5)}$: $\left(|a_1 b_1|^2, |a_1 b_2|^2\right)$. Next we continue with Step 4 of Procedure 2 by measuring the probability of getting $|000\rangle$ and $|001\rangle$ for the first three qubits. Note that we could also choose to measure $|010\rangle$ and $|011\rangle$, because any variant pair on $L_3$ will be detected by either choice of basis states. This significantly saves the number of measurements because if the variant coefficient pairs are on a very high level $L_k$, we can just choose one branch of the measurement tree (illustrated in the main text) to go down until we reach $q_k$, and do not have to measure all the $2^{k-1}$ branches before $q_k$. Now for the measurement on $|000\rangle$ and $|001\rangle$, the probability for $\psi_{\min}^{(5)}$ is $\left(|a_1 b_1 c_1|^2, |a_1 b_1 c_2|^2\right)$, but for the $\psi^{(5)}$ defined in Eq. (15) is $\left(|a_1 b_1|^2 \left(|c_3 d_1|^2 + |c_1 d_2|^2\right), |a_1 b_1|^2 \left(|c_4 d_1|^2 + |c_2 d_2|^2\right)\right)$. As explained in the main text, given enough data points of the projection measurement, we should be able to tell the difference in probability and determine the variant coefficients of $\psi^{(5)}$ as compared to $\psi_{\min}^{(5)}$. Similarly the number of projection measurements needed can be calculated by counting the leaves on the measurement tree shown in Figure 1 in the main text – only this time the number of measurements are greatly reduced because all the variant coefficients on $L_k$ and beyond share the same branch before $q_k$.